\documentclass
[twocolumn,aps,prb,reprint,showpacs,superscriptaddress,10pt,citeautoscript]{revtex4-1}%
\fontfamily{cmss}
\usepackage{graphicx}
\usepackage{dcolumn}
\usepackage{bm}
\usepackage{color}
\usepackage{dcolumn}
\usepackage{amsmath}%
\usepackage{multirow}
\usepackage{amsfonts}%
\usepackage{amssymb}
\providecommand{\U}[1]{\protect\rule{.1in}{.1in}}

\begin{document}
\title{Impact of biaxial and uniaxial strain on V$_2$O$_3$}
\author{Darshana Wickramaratne}
\affiliation{NRC Research Associate, Resident at Center for Computational Materials Science, US Naval Research Laboratory, Washington,
D.C. 20375, USA}
\author{Noam Bernstein}
\affiliation{Center for Computational Materials Science, US Naval Research Laboratory, Washington, DC 20375, USA}
\author{I. I. Mazin}
\affiliation{Center for Computational Materials Science, US Naval Research Laboratory, Washington, DC 20375, USA}

\date{\today}

\begin{abstract}
Using first-principles calculations we determine the role of compressive and tensile uniaxial and equibiaxial
strain on the structural, electronic and magnetic properties of V$_2$O$_3$.  We find that compressive strain 
increases the energy cost to transition from the high-temperature paramagnetic metallic phase
to the low-temperature antiferromagnetic insulating phase.  This shift in the energy difference can be explained by changes
in the V-V bond lengths that are antiferromagnetically aligned in the low temperature structure.  The insights that 
we have obtained provide a microscopic explanation for the shifts in the metal-insulator transition temperature that have 
been observed in experiments of V$_2$O$_3$ films grown on different substrates.
\end{abstract}
\maketitle

\section{Introduction}
\label{sec:intro}
The metal to insulator transition (MIT) in V$_2$O$_3$ \cite{rice1970metal} has been the subject of intense
investigations, due in part to the coupled structural, electronic and magnetic phase transitions
that occur in the material \cite{paolasini1999orbital,leiner2019frustrated}.  Above the bulk MIT temperature, T$_c$ of 155~K, 
V$_2$O$_3$ is a metal and is stable in the corundum phase.  Below T$_c$, 
V$_2$O$_3$ undergoes a structural transition from corundum to monoclinic.
This is accompanied by the opening of a Mott gap of 0.40 eV
below the MIT T$_c$, which manifests in a large increase in the electrical resisitivity \cite{mcwhan1973metal}.
The co-occurrence of these phenomena has led to several
efforts that have sought to control these phase transitions.

To understand and control the MIT in V$_2$O$_3$ it is important to consider the underlying microscopic
mechanism that leads to the transition.  While it is well accepted that above T$_c$ the electronic phase is
metallic and below T$_c$ it is a Mott insulator, the mechanism leading to the MIT cannot be 
described as a Mott transition.  It is only recently, through 
a combination of neutron scattering measurements and first-principles calculations,  that Leiner {\it et al.} convincingly 
demonstrated that it is instead a first-order phase transition between two states that host 
different magnetically ordered states in addition to being structurally and electronically distinct above and below
the MIT T$_c$ \cite{leiner2019frustrated}.
The metallic high-temperature (HT)
phase was shown to be a strongly frustrated paramagnet and the insulating low-temperature (LT) phase is 
a robust antiferromagnet with little frustration.
Since the structural, electronic and magnetic properties of V$_2$O$_3$ are intimately linked, this 
makes the metal-insulator transition temperature sensitive to external perturbations.

Indeed, we have shown that the presence of point defects in the form of Frenkel pairs 
disrupts bonding and the magnetic ordering of the V atoms, which in turn leads to a reduction in the
energy to transition between the HT paramagnetic metallic phase and the LT antiferromagnetic
insulating phase \cite{wickramaratne2019role}.  This is consistent with an experimental observation that found
the MIT T$_c$ to decrease when point defects are introduced intentionally compared to the $T_c$ of
as-grown V$_2$O$_3$ \cite{Ramirez_2015}.
The sensitivity of the MIT to changes in bonding has also made the use
of strain an appealing approach to manipulate and control the MIT.
This is part of a general growing interest in manipulating the transition temperature
of materials that exhibit a MIT by taking advantage of advances in 
epitaxial growth, which has enabled the growth of thin films on
targeted substrates \cite{saerbeck2014coupling}.

Epitaxial growth of V$_2$O$_3$ on $a$-plane ($11\bar{2}0$), $c$-plane ($0001$), $m$-plane ($1\bar{1}00$)
and $r$-plane ($1\bar{1}02$) Al$_2$O$_3$ substrates has been explored by a number of groups 
\cite{thorsteinsson2018tuning,brockman2012substrate,kalcheim2019robust,dillemans2014evidence,allimi2008resistivity,schuler1997influence, sass2003structural}.  Growth of V$_2$O$_3$ on these substrates occurs at a 
temperature well above the MIT T$_c$, which, results in a film that presumably adopts the high-temperature (HT) paramagnetic
corundum structure during growth.  The Al$_2$O$_3$ lattice constants are lower than the corundum V$_2$O$_3$ lattice constants, so V$_2$O$_3$ is 
expected to be under compressive strain if the growth is coherent.
  The reports of T$_c$ identified from measurements of resistance versus temperature of these epitaxially
grown films are varied.  Schuler {\it et al.} demonstrated that the T$_c$ increases by 45 K with respect to unstrained V$_2$O$_3$ in a cooling cycle
measurement of resistance versus temperature for V$_2$O$_3$ grown on $c$-plane Al$_2$O$_3$ \cite{schuler1997influence}. 
In contrast, Kalcheim {\it et al.} \cite{kalcheim2019robust} 
demonstrated that V$_2$O$_3$ grown on the $m$-plane and $r$-plane orientations of Al$_2$O$_3$ led to a T$_c$ that is larger
than the unstrained T$_c$ (by up to 16 K), while growth on the $a$-plane orientation of Al$_2$O$_3$ led to a reduction in
T$_c$.  Growth on alternative substrates, such as LiTaO$_3$, has also been explored where V$_2$O$_3$ is expected
to be under tensile strain if the growth is coherent \cite{allimi2008thickness, brockman2012substrate,yonezawa2004epitaxial}.  
In these studies it was found the T$_c$ of V$_2$O$_3$ was larger than the T$_c$ of unstrained V$_2$O$_3$.
An alternative approach to impart strain on V$_2$O$_3$ has been through the use of ferroelectric and piezoelectric
substrates that are subject to electrical biases with different polarity \cite{salev2019giant, sakai2019strain}.
For instance, the T$_c$ of V$_2$O$_3$ on a PMN-PT substrate increased by 30 K when the PMN-PT substrate 
underwent tensile expansion due to an applied bias.  

The results of these experimental studies have been interpreted 
using the pressure versus temperature phase diagram of V$_2$O$_3$ \cite{rice1970metal}.  According to this phase diagram,
positive pressure (volume reduction) leads to a reduction in T$_c$ while negative pressure (volume
expansion) leads to an increase in T$_c$.  
However, for a material under uniaxial or biaxial
strain, the corresponding change in bonding can be very different from the change in bonding
associated with hydrostatic pressure.
Furthermore, first-principles calculations have demonstrated that relying on chemical pressure alone to 
interpret the V$_2$O$_3$ phase diagram can be misleading \cite{lechermann2018uncovering}.
To our knowledge, the impact of strain
on the structural, magnetic and electronic properties of V$_2$O$_3$ has not yet been theoretically studied.
Because of this, there is no clear relationship
that can be deduced from the experimental reports on changes in the V$_2$O$_3$ T$_c$
grown on different substrates and the magnitude and direction of the strain imparted.

In this study, we use first-principles calculations to investigate
the effect of uniaxial and equibiaxial compressive and tensile strains on the structural, magnetic and electronic
properties of V$_2$O$_3$.  We find that up to 1\%~compressive equibiaxial or uniaxial strains increase the energy required
to transition to the LT antiferromagnetic insulating phase by up to 75$\%$ compared to unstrained
V$_2$O$_3$.  This would be reflected in an increase in the MIT T$_c$ when V$_2$O$_3$ is under compressive strain.
In contrast, we find equibiaxial tensile or uniaxial tensile strains lead
to modest reductions or increases in the energy to transition to the antiferromagnetic insulating phase depending on the
direction along which the strain is imparted.  We identify the microscopic origin of these changes in the energy
to transition between the metallic and insulating phase as being changes in the bond lengths of the pair of next-nearest
neighbor vanadium atoms that are antiferromagnetically aligned in the LT monoclinic structure.

\section{Computational Methods}
\label{sec:methods} 
Our calculations are based on density functional theory
within the projector-augmented wave method \cite{Blochl_PAW} as implemented in
the \textrm{VASP} code \cite{VASP_ref,VASP_ref2} using the generalized
gradient approximation defined by the Perdew-Burke-Ernzerhof (PBE) functional
\cite{perdew1996generalized}. In our calculations, \textrm{V} $4s^{2}%
3p^{6}3d^{3}$ electrons and \textrm{O} $2s^{2}2p^{4}$ electrons are treated as
valence. All calculations use a plane-wave energy cutoff of 600~eV.  Structural
relaxations of the lattice parameters and internal coordinates were carried
out with an $8\times8\times8$ $k$-point grid and a force convergence
criterion of 5 meV/\AA.~
In order to
simulate the Mott-insulating behavior of V$_{2}$O$_{3}$ 
we use a spherically-averaged Hubbard
correction within the fully-localized limit double-counting subtraction
\cite{anisimov1993density}.
We apply a $U-J$ value of 1.8~eV
to the V $d$-states, which reproduces the experimental band gap
of V$_2$O$_3$.  We note studies that compared exchange coupling constants obtained
from neutron scattering with first-principles calculations relied on a larger
value of $U-J$ (3 eV) to obtain quantitative agreement between theory and experiment \cite{leiner2019frustrated}.
We find our overall conclusions to remain unchanged if we also use a $U-J$ value of 3 eV.

To study the effects of epitaxial strain, we performed
``strained-bulk" calculations where we impose compressive and
tensile equibiaxial strain on the $a$ and $b$ (denoted as $ab$), $b$ and $c$ (denoted as $bc$), $a$ and
$c$ (denoted as $ac$) monoclinic lattice vectors and uniaxial strain along the 
$a$, $b$ and $c$ monoclinic lattice vectors of the
V$_2$O$_3$ unit cell and then optimize the free lattice constant(s)
and all atomic positions of the unit cell.
The standard VASP package does
not allow for arbitrary constraints to be placed on the strain
tensor during relaxation.  To perform these constrained calculations,
we made modifications that set specific components of the stress
tensor to zero during the minimization routine.  This allowed us to impose
strain along the different axes as we report here.

Since growth of V$_2$O$_3$ occurs at temperatures well above the MIT T$_c$, the 
as-deposited V$_2$O$_3$ epitaxial films will adopt the HT paramagnetic
structure.  
Paramagnetically ordered states are challenging to describe
with standard DFT.  However, since the paramagnetic HT phase is magnetically frustrated \cite{leiner2019frustrated},
magnetic ordering has a weak effect on total energies.  We have previously shown that the FM ordered
monoclinic structure can be used as a suitable proxy for the paramagnetic HT corundum phase \cite{wickramaratne2019role}.
Since strain is defined with respect to the HT phase, we use the lattice constants
of the HT ferromagnetic structure as the reference for strain.
For example, uniaxial strain along the monoclinic $a$ axis, $\epsilon_{a}$, is
defined as: $\epsilon_{a}$ = [($a$ - $a_{0}$)/$a_{0}$], where $a_0$ is the equilibrium $a$ lattice
constant of the FM monoclinic structure.  In such a calculation we would only
allow the monoclinic $b$ and $c$ lattice constants, bond angles and atomic coordinates to be 
optimized.  Next, with this optimized structure we impose an AFM ordering of spins on the V atoms (ferromagnetic
along the $a$ and $c$ axes and antiferromagnetic along the $b$ axis) and optimize the free lattice parameters and
atomic coordinates.
We report results for compressive and tensile strain along each of the monoclinic axes for 
strains that range between $\pm$ 1$\%$.  Positive values of $\epsilon$ correspond to tensile strain.

\section{Results}
\label{sec:results}

\subsection{Bulk properties}
The HT metallic phase of V$_2$O$_3$ is stable in the corundum structure with space group $R\bar{3}c$.
Neutron scattering measurements in combination with first-principles calculations have
demonstrated the corundum phase of V$_2$O$_3$ to be a highly frustrated paramagnet \cite{leiner2019frustrated}.
As we discuss in Sec.~\ref{sec:methods}, we use a FM ordered structure as a proxy for the
disordered paramagnetic phase of the corundum HT structure since it has the correct magnitude of the 
magnetic moments and it respects the full
lattice symmetry (as opposed to an antiferromagnetic arrangement of spins).
Indeed, we have previously shown that the energy difference between the corundum structure with FM
order and AFM order imposed is low, 0.8 meV per vanadium atom \cite{wickramaratne2019role}, consistent with the magnetic
frustration that has been experimentally identified in the HT phase \cite{leiner2019frustrated}.
Our DFT+$U$ lattice constants of the FM corundum structure are 
$a$=$b$=5.037 \AA, and $c$=14.305 \AA~and the bond angle of the rhombohedral unit cell is
$\theta$=54.6$^{\circ}$, which are within 1.5$\%$ of the experimentally measured lattice
parameters of the HT corundum structure ($a$=$b$=4.952 \AA, $c$=14.003 \AA~and $\theta$=56.1$^{\circ}$).

The LT insulating phase of V$_2$O$_3$ is antiferromagnetic and has a monoclinic structure with space group $P_{2}1/c$.  
We find the lattice
constants of the LT monoclinic structure of V$_{2}$O$_{3}$ to be: $a$ = 7.414
\AA , $b$ = 5.084 \AA ~and~$c$ = 5.559 \AA, and the bond angles to be 
$\alpha$=$\gamma$=90$^{\circ}$ and $\beta$=97.3$^{\circ}$, which are within 2.7$\%$ of the experimental LT
lattice parameters ($a$ = 7.255 \AA , $b$ = 5.002 \AA, $c$ = 5.548 \AA~and $\beta$=96.8$^{\circ}$)
reported for monoclinic V$_{2}$O$_{3}$ \cite{dernier1970crystal}.
To describe the antiferromagnetic ordering of spins we use a four formula-unit cell.  We find
the ground state magnetic ordering to be the one where the V atoms are aligned ferromagnetically along the
monoclinic $a$ and $c$ axes and aligned antiferromagnetically along the monoclinic $b$ axis, which
is consistent with neutron scattering measurements of the monoclinic insulating phase \cite{leiner2019frustrated, moon1970antiferromagnetism}.

\begin{figure}
\includegraphics[width=6.5cm]{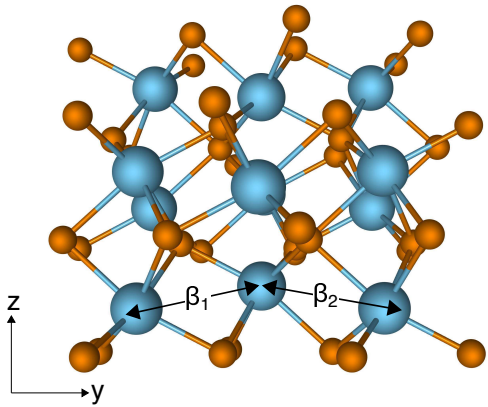}\caption{Schematic illustration of the 
V$_2$O$_3$ monoclinic unit cell.  Vanadium
atoms are in blue and oxygen atoms are in orange.  The pair of V atoms that are antiferromagnetically
aligned along the monoclinic $b$ axis are denoted $\beta_{1}$ and $\beta_{2}$.  
}
\label{fig:uc}
\end{figure}

Along the monoclinic $b$-axis, the pair of next-nearest neighbor V atoms that are antiferromagnetically aligned have
two different V-V bond lengths.  We label the shorter of the two bonds $\beta_1$, and the second
V-V pair is labeled $\beta_2$, as shown in Fig.~\ref{fig:uc}.  The bond length of the $\beta_1$ pair is  
2.996 \AA,~while the bond length of the $\beta_2$ pair is 3.085 \AA.

Since we are interested in the impact of strain on the energy to transition between the paramagnetic (approximated
as ferromagnetic) HT phase and the antiferromagnetic LT phase using the ``strained-bulk'' approach,
we also calculate the lattice parameters and
electronic properties of the monoclinic structure where all of the V atoms are ferromagnetically
aligned.  
If we allow for full structural relaxation (volume, cell shape and atomic positions) in this magnetic state we find the structure
takes on the HT corundum structure and is metallic.  

\subsection{Biaxial and uniaxial strain}
\subsubsection{Structural properties}
For the magnitudes of strain that we have investigated, we find that biaxial and uniaxial strain leads to elastic
changes in the volume and in turn in the free lattice parameter(s).  The monoclinic bond angle only changes by up to
$\pm$0.2$\%$ for the
largest strain ($\pm$1$\%$) that we consider.

If we consider biaxial strain imposed along the monoclinic $ab$ axes, applying compressive equibiaxial strain
to the FM structure and allowing the monoclinic $c$ lattice parameter and all atomic positions to relax 
leads to an increase in the $c$ lattice constant.  We find the $c$ lattice constant increases
linearly as a function of the applied compressive strain.  Conversely, for equibiaxial tensile strain, the $c$ lattice
constant decreases linearly with respect to the $c$ lattice constant of the unstrained FM structure. 
The ratio of the change in the $c$ lattice constant
as a function of the applied equibiaxial in-plane strain is a positive constant in the elastic regime and is defined as 
the Poisson ratio, $\nu = -\epsilon_{zz}/(\epsilon_{xx}+\epsilon_{yy})$, where
$\epsilon_{zz}$ is the strain in the $c$ lattice constant and $\epsilon_{xx}$ and $\epsilon_{yy}$ are the
strains along the $a$ and $b$ monoclinic lattice constants, respectively.  
We find $\nu$=0.31 for the FM structure under equibiaxial
strain along the $ab$ axes.  We find the response of the V$_2$O$_3$ lattice to equibiaxial strain along the $bc$ and $ac$ axes
to be similar; the free lattice parameter changes linearly with a Poisson ratio that is positive.  $\nu$=0.33 for
the FM structure under equibiaxial strain along $bc$ and $\nu$=0.32 for equibiaxial strain along $ac$.

Next we impose AFM ordering on the structures that are under equibiaxial or uniaxial tensile strain.
As discussed in Section \ref{sec:methods}, we assume the lattice parameters of the HT paramagnetic (approximated as FM)
would be clamped to the substrate post-growth and these lattice parameters remain fixed when the 
transition to the LT AFM insulating phase occurs.
For example, for equibiaxial strain along the monoclinic $ab$ axis, $\epsilon_{ab}$, 
of 1$\%$ (where $\epsilon_{ab}$ is with respect to the equilibrium FM lattice parameters) we use the same monoclinic $a$ and
$b$ lattice parameters that are strained by 1$\%$ with respect to the equilibrium FM monoclinic lattice constants, 
impose AFM order and allow the monoclinic $c$ lattice constant and all atomic coordinates to relax.

For each of the structures under strain with AFM order imposed, we find the lattice also responds elastically with a
positive Poisson ratio.  The magnitude of $\nu$ for the structures that are antiferromagnetically
ordered for the different directions of equibiaxial strain are as follows: $\nu$=0.36 ($ab$), 0.38 ($bc$)
and 0.33 ($ac$).

\subsubsection{Total energies}
\begin{figure}
\includegraphics[width=8.5cm]{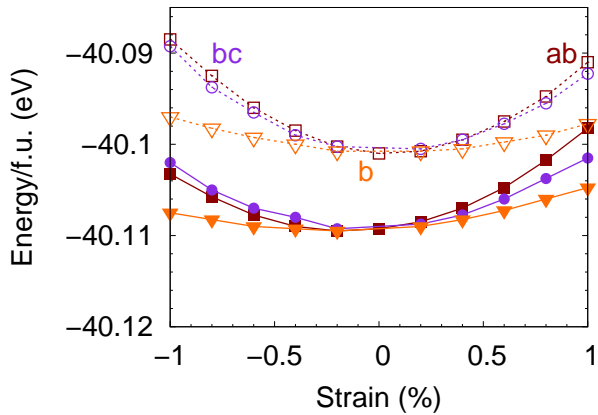}\caption{Variation in the total energy of the AFM (solid line) and FM (dotted line)
configuration per V$_2$O$_3$ formula unit 
under compressive and tensile strain along the monoclinic $ab$ ($\square$), $bc$ ($\circ$) and $b$ ($\triangledown$) axes.
Note, strain is defined with respect to the equilibrium lattice constants of the FM structure.}
\label{fig:en}
\end{figure}

In Figure~\ref{fig:en}, we illustrate the variation in the total energy of the FM and AFM configuration under
biaxial strain along the monoclinic $ab$ and $bc$ axes and uniaxial strain along the monoclinic $b$ axis.
It is evident that the AFM configuration remains lower in energy than the
FM configuration for all values of strain.  For each of these values of strain, the FM configuration
remains metallic while the AFM configuration remains insulating.

\subsubsection{Spin-flip energies}
\begin{figure}
\includegraphics[width=8.5cm]{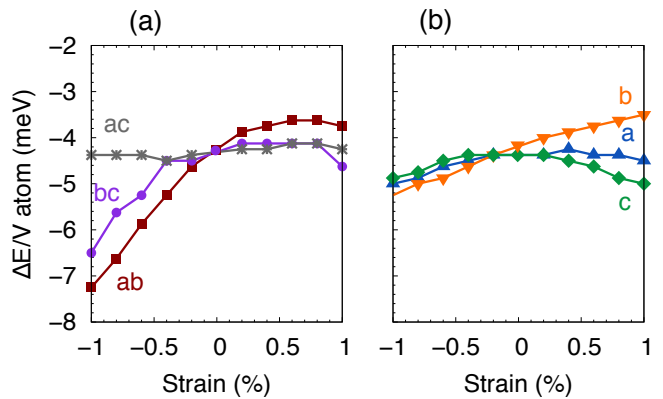}\caption{Spin-flip energy, $\Delta E$, per V atom
as a function of compressive and tensile (a) biaxial strain along the 
monoclinic $ab$ ($\square$), $bc$ ($\circ$), $ac$ ($\ast$) axes and (b) uniaxial strain
along the monoclinic $a$ ($\bigtriangleup$), $b$ ($\triangledown$) and $c$ ($\Diamond$) axes.
Note, strain is defined with respect to the equilibrium lattice constants of the FM structure.}
\label{fig:de_flip}
\end{figure}

From Fig.~\ref{fig:en} it is also evident the energy to transition between the high-temperature FM configuration
and the low-temperature AFM configuration as a function of compressive and tensile strain is not a constant.
For example, the total energy difference between the 
AFM and FM configuration under biaxial strain along $ab$ is larger under compressive strain compared to tensile strain.
  We define this energy difference between the AFM and FM configuration
at a fixed strain, $\epsilon$, a spin-flip energy, $\Delta E$, where
$\Delta E$ = [$E_{\rm tot}$(AFM) - $E_{\rm tot}$(FM)].  
The spin-flip energy as a function
of equibiaxial and tensile strain is illustrated in Fig.~\ref{fig:de_flip}.

We first consider the change in spin-flip energy for the structures subject to equibiaxial strain.
Under compressive strain along the $ab$ and $bc$ axes, 
$\Delta E$ decreases with respect to the equilibrium spin-flip energy by up to 75$\%$ at the largest
value of $\epsilon$ of 1$\%$. 
Tensile strain along these axes leads to a modest increase in the
spin-flip energy for strain along $ab$ and a modest reduction in $\Delta E$ for $\epsilon$ greater
than 0.5$\%$ along $bc$.
In contrast, we find $\Delta E$ is insensitive to compressive and tensile equibiaxial strain along the monoclinic $ac$ axes.

When V$_2$O$_3$ is subject to uniaxial strain, we find the change in $\Delta E$ to be modest
in comparison to the change in $\Delta E$ under biaxial strain.  In particular, when the monoclinic
$a$, $b$ or $c$ axes are under compressive strain, we find they all lead to 
a slight decrease in $\Delta E$.  Under tensile uniaxial strain, $\Delta E$ decreases for strain along the $a$
and $c$ axes and increases for tensile strain along the $b$ axis.

\section{Discussion}
At this point it is instructive to examine the primary contributions to the change in $\Delta E$
under compressive and tensile strain.  We decompose this change in $\Delta E$ into two contributions, 
an elastic energy, $\Delta E^{\rm el}$, and a magnetic energy, $\Delta E^{\rm mag}$, such that
$\Delta E = \Delta E^{\rm el} + \Delta E^{\rm mag}$.
The elastic energy, $\Delta E^{\rm el}$, is the change in energy due to the change in the lattice
parameters and the atomic positions to transition from the geometry associated with the FM to the 
AFM configuration at a fixed magnetic configuration.  
We define $\Delta E^{\rm el}$ as ($E^{[\epsilon,\rm FM]}_{\rm tot}$(AFM) - $E^{[\epsilon,\rm AFM]}_{\rm tot}$(AFM)) 
where $E^{[\epsilon,\rm FM]}_{\rm tot}$(AFM) is the total energy of the structure with 
the atomic coordinates and lattice parameters of V$_2$O$_3$ in the 
strained FM configuration with AFM order imposed 
and $E^{[\epsilon,\rm AFM]}_{\rm tot}$(AFM) is the total energy of the structure with 
the atomic coordinates and the lattice parameters in the strained AFM configuration and AFM order imposed.
The magnetic energy, $\Delta E^{\rm mag}$, is the change in energy associated with
flipping spins from ferromagnetic to antiferromagnetic at a fixed set of atomic coordinates and lattice parameters.  
We define $\Delta E^{\rm mag}$ as ($E^{[\epsilon,\rm AFM]}_{\rm tot}(\rm AFM) - {\it E}^{[\epsilon,\rm AFM]}_{\rm tot}(\rm FM)$)
where $E^{[\epsilon,\rm AFM]}_{\rm tot}(\rm AFM)$ is the total energy of the structure with the atomic coordinates
and lattice parameters of the strained AFM configuration with AFM order imposed and 
${\it E}^{[\epsilon,\rm AFM]}_{\rm tot}(\rm FM)$ is the total energy of the structure with the atomic coordinates
and lattice parameters of the strained AFM configuration with FM order imposed.
For all values of biaxial and uniaxial strain, $\Delta E^{\rm el}$ only changes by up to 0.5 meV per vanadium atom.
Note, $\Delta E$ changes by up to $\sim$4 meV per vanadium atom in comparison to unstrained V$_2$O$_3$ (Fig.~\ref{fig:de_flip}).
Hence, the remaining energy difference between $\Delta E$ and $\Delta E^{\rm el}$ is the change in the
magnetic energy, $\Delta E^{\rm mag}$, as a function of strain.

Based on Fig.~\ref{fig:de_flip} it is evident that $\Delta E$ is more sensitive to compressive strain in V$_2$O$_3$.
To explain this sensitivity to compressive strain, we examine the bond lengths of
V$_2$O$_3$ in the AFM configuration.  In the AFM configuration, the V atoms are ferromagnetically coordinated along the $a$
and $c$ axes and antiferromagnetically coordinated along the $b$ axis.  Previous first-principles calculations of the unstrained
V$_2$O$_3$ lattice \cite{leiner2019frustrated} have demonstrated that the V atoms along the
monoclinic $b$ axis have the largest exchange coupling constants compared to 
the next nearest neighbor V-V exchange coupling constants along the other axes of the monoclinic structure.
Within a nearest
neighbor Heisenberg model, the N\'eel temperature of the HT paramagnetic to the LT AFM phase transition would be
determined primarily by the exchange coupling constants of these antiferromagnetically aligned V atoms.
We denote these V-V bonds along the monoclinic $b$ axis as $\beta_1$ and $\beta_2$ (Fig.~\ref{fig:uc}), where the
bond length of $\beta_1$ ($d_{\beta_1}$) is lower than the bond length of $\beta_2$ ($d_{\beta_2}$).
Leiner {\it et al.} \cite{leiner2019frustrated} have shown the exchange coupling constant of
$\beta_1$ is twice larger than $\beta_2$.

\begin{figure}
\includegraphics[width=8.5cm]{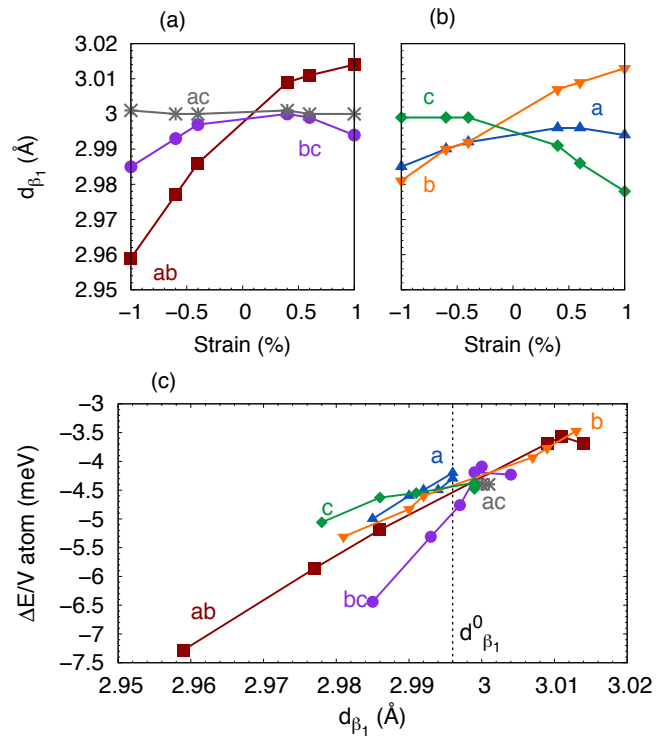}\caption{Change in the bond length, $d_{\beta_{1}}$,
(cf. $\beta_{1}$ in Fig.~\ref{fig:uc}) as a function of compressive and tensile
(a) biaxial strain along the
monoclinic $ab$ ($\square$), $bc$ ($\circ$), $ac$ ($\ast$) axes and (b) uniaxial strain
along the monoclinic $a$ ($\bigtriangleup$), $b$ ($\triangledown$) and $c$ ($\Diamond$) axes.
Note, strain is defined with respect to the equilibrium lattice constants of the FM structure.
(c)  Spin-flip energy versus change in the $d_{\beta_{1}}$ bond length for the different directions
of compressive and tensile strain imparted on the monoclinic axes.  The $d_{\beta_{1}}$ bond length of
unstrained AFM V$_2$O$_3$ ($d^{0}_{\beta_{1}}$=2.996 \AA) is shown with a dotted black line.}
\label{fig:vbond}
\end{figure}

Through our first-principles calculations we find that $d_{\beta_{1}}$ 
changes non-monotonically as a function of compressive and tensile strain.
These results are illustrated in Fig.~\ref{fig:vbond}(a) and (b).
Figure \ref{fig:vbond}(c) illustrates
the dependence of $\Delta E$ on $d_{\beta_{1}}$ for the different directions of strain
we consider in our study.  When $d_{\beta_{1}}$ decreases with respect to $d_{\beta_{1}}$ of the unstrained
AFM monoclinic structure, we find $\Delta E$ decreases, while an 
increase in $d_{\beta_{1}}$ corresponds to an increase in $\Delta E$.
Since the primary contribution to the change in $\Delta E$ is the magnetic energy,
this dependence of $\Delta E$ on $d_{\beta_{1}}$ can be understood as follows.
A reduction in the $d_{\beta_1}$ bond length is expected to lead to an increase in the hopping energy, $t$,
between the $\beta_1$ pair of vanadium atoms (Fig.~\ref{fig:uc}),
which in turn would lead to an increase in the exchange coupling constant, $J_{\beta_{1}}$, where $J_{\beta_{1}} \propto -t^{2}/U$ 
and $U$ is the on-site Coulomb repulsion.  Conversely, we expect an increase in $d_{\beta_{1}}$ to 
lead to a reduction in $J_{\beta_{1}}$ compared to unstrained V$_2$O$_3$.  
This dependence of $\Delta E$ on $d_{\beta_{1}}$ also explains why equibiaxial compressive
and tensile strain along the 
monoclinic $ac$ axes does not lead to a change in $\Delta E$.
We find the biaxial strain that is imparted on the $ac$ axes
is accommodated by changes in $d_{\beta_{2}}$ while $d_{\beta_{1}}$ remains unchanged for all values of strain that we 
investigate (Fig.~\ref{fig:vbond}(a)).

Hence, our calculations suggest that the MIT T$_c$ is sensitive to changes in
the bond length, $d_{\beta_{1}}$.  
In particular, we suggest
compressive strain along the monoclinic
$bc$, $ab$, $a$, $b$ and $c$ axes and tensile strain along the monoclinic $bc$, $a$ and $c$ axes 
will increase T$_c$ compared to unstrained V$_2$O$_3$. 
We note that this is consistent with the increase in the MIT T$_c$ that has been measured in
V$_2$O$_3$ thin films grown on Al$_2$O$_3$ substrates where V$_2$O$_3$ is under compressive
strain \cite{thorsteinsson2018tuning,brockman2012substrate,kalcheim2019robust,schuler1997influence}.  

\section{Summary and Conclusions}
\label{sec:conclusions}
In summary, we examined the role of equibiaxial and uniaxial compressive and tensile
strain on the electronic, structural and 
magnetic properties of V$_2$O$_3$.

The metal-insulator transition in V$_2$O$_3$ was recently re-interpreted as being 
a strong first-order transition between the 
high-temperature corundum structure, which is a highly-frustrated paramagnet and the low-temperature
monoclinic structure that is strongly antiferromagnetic \cite{leiner2019frustrated}.  The leading contribution to the strong
antiferromagnetic coupling in the low-temperature monoclinic phase is the shortest of the pair
of V-V bonds that are antiferromagnetically aligned ($\beta_1$) along the monoclinic $b$ axis.
As a result, shifts in the energy difference between the high-temperature metallic phase and the low-temperature
insulating phase are sensitive to changes in the bond length of $\beta_1$ of the LT AFM monoclinic phase.
Our calculations confirm this interpretation and demonstrate that changes in the bond length of 
$\beta_1$ due to strain can lead to changes in this energy difference.
In particular, we find that a suppression of this energy difference which would translate to a
reduction in the MIT T$_c$ coincides with an elongation of the bond length of $\beta_1$ 
while an increase of this energy difference coincides with a compression of the $\beta_1$ bond length.

Based on our calculations we can draw the following conclusions on the
role of strain on the metal-insulator transition temperature of V$_2$O$_3$.
Under compressive strain
along the monoclinic $bc$, $ab$, $b$, $a$ and $c$ axes, the energy to transition to the 
low-temperature insulating antiferromagnetic phases
increases by up to 75$\%$ for compressive strains up to 1$\%$.  Tensile strain along
the monoclinic $a$ and $c$ axes lead to modest increases in the energy to transition
to the insulating antiferromagnetic phase.  Hence, strain along these directions and axes
will likely lead to an increase in T$_c$ compared to unstrained V$_2$O$_3$. 
Tensile strain along the $ab$ and $b$ axes lowers the energy to transition
to the insulating phase by up to 10$\%$ compared to unstrained
V$_2$O$_3$ for the largest strain we consider of 1$\%$; which would be reflected in a reduction
in T$_c$ compared to unstrained V$_2$O$_3$.
In contrast, compressive and tensile strain along the monoclinic $ac$ axes does not lead
to any change in the energy to transition between the insulating and metallic phase.

\acknowledgements
We thank Ivan Schuller for helpful discussions and for
encouraging us to undertake this work within the framework
of the LUCI collaboration and Ivan Schuller’s Vannevar Bush
Faculty Fellowship program. D.W. acknowledges support
from the National Research Council fellowship at the U.S.
Naval Research Laboratory. The work of N.B. and I.I.M.
was supported by the Laboratory-University Collaboration
Initiative (LUCI) 
of the Office of the Under Secretary of Defense 
for Research $\&$ Engineering Basic Research Office.


%
\end{document}